\documentstyle[aps,prb,epsfig]{revtex}
\begin{document}
\baselineskip=0.5cm
\renewcommand{\thefigure}{\arabic{figure}}
\title{Spin-density functional approach to thermodynamic and structural consistence in the charge and spin response of an electron gas}
\author{B. Davoudi$^{1,2}$, M. Polini$^1$ and M. P. Tosi$^1$~\footnote{Corresponding author: tosim@sns.it}}
\address{
$^1$NEST-INFM and Classe di Scienze, Scuola Normale Superiore, I-56126 Pisa, Italy\\ 
$^2$Institute for Studies in Theoretical Physics and Mathematics, Tehran, P.O.Box 19395-5531,Iran\\
}
\maketitle
\vspace{1 cm}

{\bf Abstract}
\vspace{0.2 cm}

We use spin-density functional theory to obtain novel expressions for the charge and spin local-field factors of an electron gas in terms of its electron-pair structure factors. These expressions (i) satisfy the compressibility and spin susceptibility sum rules; (ii) keep account of kinetic correlations by means of an integration over the coupling strength; and (iii) provide a practical self-consistent scheme for evaluating linear response and liquid structure. Numerical illustrations are given for the dielectric response of the paramagnetic electron gas in both three and two dimensions.
\vspace{0.3cm}


\vspace{0.8cm}

 An interacting electron gas on a uniform neutralizing background is the reference system for many calculations of electronic structure in condensed matter systems [1]. Its dielectric and spin susceptibilities provide key inputs in density functional theory [2] and in studies of quasi-particle properties such as the effective mass and the effective Land\'e g-factor [3]. These properties are now known from experiment for carriers in semiconductor structures over a wide range of density [4]. A great deal of relevant information has been coming from Quantum Monte Carlo (QMC) studies [5 - 9], but theoretical understanding continues to attract interest.
 
The issue of thermodynamic consistence in the theory of linear response for an electron gas, though raised in the early work of Vashishta and Singwi [10] through an {\it ad hoc} scheme to account for the compressibility sum rule, has not received much attention. In brief, thermodynamic consistence requires that the linear susceptibilities (at zero temperature) should reduce in the static, long-wavelength limit to the values obtained from the second derivatives of the (ground-state) energy. Of course, it is possible to model the susceptibilities by parametrized analytical expressions that embody these limiting behaviors as they are known from QMC data [11, 12]. In this Letter we propose instead a fully self-consistent theoretical scheme, which uses the mean particle and spin densities of the system as the only inputs for the calculation of the susceptibilities and of the electron-pair structure. Our approach is aimed at satisfying both the compressibility and the spin susceptibility sum rule: indeed, the QMC studies show that spin correlations and spin ordering are enhanced on increasing the coupling strength.	Our starting point is the expression of the exchange-correlation energy $E_{\mathrm xc}[n_\uparrow,n_\downarrow]$ as a functional of the spin densities $n_{\uparrow}({\bf r})$ and $n_{\downarrow}({\bf r})$ in an inhomogeneous $d$-dimensional electron gas with interaction potential $v(r)=e^2/r$,
\begin{equation}\label{exc1}
E_{\mathrm xc}[n_\uparrow, n_\downarrow]=
\frac{1}{2\,e^2}\int_{0}^{e^2}d\lambda\,\sum_{\sigma, \sigma'}\int d^{d}{\bf r}\int d^{d}{\bf r}'\,
v(|{\bf r}-{\bf r}'|)~\left[\,\langle \Psi_{\lambda}|\delta {\widehat n}_{\sigma}({\bf r})\delta {\widehat n}_{\sigma'}({\bf r}')|\Psi_{\lambda}\rangle-\delta_{\sigma\sigma'}n_{\sigma}({\bf r})~\delta({\bf r}-{\bf r}')\right]
\end{equation}
where $|\Psi_{\lambda}\rangle$ is the ground state of the gas with interaction potential   $v_{\lambda}(r)=\lambda/r$ and $\delta {\widehat n}_{\sigma}({\bf r})={\widehat n}_{\sigma}({\bf r})-n_{\sigma}({\bf r})$. The spin-density operators ${\widehat n}_{\sigma}({\bf r})$ are given by ${\widehat n}_{\sigma}({\bf r})= \psi^{\dagger}_{\sigma}({\bf r})\psi_{\sigma}({\bf r})$ in terms of Schr\"odinger field operators obeying canonical anticommutation relations. In deriving Eq. (1), which is obtained by an immediate extension of the treatment given by Dreizler and Gross [2] for the paramagnetic state, the technique of coupling-constant integration has been adapted to the context of density functional theory in order to account for kinetic correlations. Upon introducing as independent variables the total particle density $n({\bf r})=n_{\uparrow}({\bf r})+n_{\downarrow}({\bf r})$ and the spin magnetization density $m({\bf r})=n_{\uparrow}({\bf r})-n_{\downarrow}({\bf r})$, Eq. (1) can be written in the more convenient form		 
\begin{equation}\label{exc2}
E_{\mathrm xc}[n, m]=\frac{1}{2}\,\int\,d^{d}{\bf r}\int\,d^{d}{\bf r}'\,v(|{\bf r}-{\bf r}'|)\, n({\bf r}) n({\bf r}')\,\left[{\bar g}({\bf r}, {\bf r}')-1\right]
\end{equation}
where the coupling-averaged pair function ${\bar g}({\bf r}, {\bf r}')$ is defined by
\begin{equation}	
{\bar g}({\bf r}, {\bf r}')=\frac{1}{e^2}\int_0^{e^2}d\lambda ~g_\lambda({\bf r}, {\bf r}').
\end{equation}			                 
Here, $g_\lambda({\bf r}, {\bf r}')$ is the pair distribution function in the gas with interaction potential $v_\lambda(r)$  of strength $\lambda$  and is, of course, a functional of $n$ and $m$.	
 
We are in fact interested in the linear-response regime, where the inhomogeneity of the system is due to a weak static perturbation. We can then introduce the exchange-correlation kernels for the {\it homogeneous} electron gas as		                         	
\begin{equation}\label{kxcs}
K^{\sigma \sigma'}_{\rm \scriptstyle xc}({\bar n}, {\bar m};
|{\bf r}-{\bf r}'|)\equiv 
\left.\frac{\delta^{2} E_{
\mathrm xc}}{\delta n_{\sigma}({\bf r}) \delta
n_{\sigma'}({\bf r}')}\right|_{\rm \scriptstyle hg}
\end{equation}
and the local field factors $G_{\sigma\sigma'}(q)$ as		                         
\begin{equation}\label{lff_sp}
v_q~G_{\sigma\sigma'}(q)\equiv -\,\int d^d{\bf r}~K^{\sigma \sigma'}_{\rm \scriptstyle xc}(r)\,\exp{(-i{\bf q}\cdot {\bf r})}~,
\end{equation} 
with $v_q$ the Fourier transform of the electron-electron potential. In Eq. (4), after taking the second functional derivative the inhomogeneous spin densities are replaced by those of the homogeneous gas, corresponding to a particle density ${\bar n}$  and a magnetization density ${\bar m}$.	

While the formulation given so far is quite general (within neglect of spin-orbit coupling), in the following we restrict ourselves for simplicity to the paramagnetic state ({\it i.e.} ${\bar m}=0$). In this case the matrices in Eqs. (4) and (5) can be diagonalized and expressed in terms of the charge and longitudinal-spin response kernels,		                 	
\begin{equation}\label{kxc1}
K^{+}_{\mathrm  xc}({\bar n};
|{\bf r}-{\bf r'}|)\equiv 
\left.\frac{\delta^2 E_{\mathrm xc}}{\delta n({\bf r}) \delta
n({\bf r}')}   \right|_{\rm \scriptstyle hpg}
\end{equation}
and
\begin{equation}\label{kxc2}
K^{-}_{\mathrm  xc}({\bar n};
|{\bf r}-{\bf r'}|)\equiv 
\left.\frac{\delta^2 E_{\mathrm xc}}{\delta m({\bf r}) \delta
m({\bf r}')}   \right|_{\rm \scriptstyle hpg}~.
\end{equation}
The corresponding local field factors are $G_+(q)$ and $G_-(q)$, say. Notice that in the calculation of $K^{-}_{\mathrm  xc}({\bar n};r)$ the functional derivatives in Eq. (7) must be evaluated on the spin-polarized gas {\it before} taking the paramagnetic limit.	

The evaluation of the functional derivatives in Eqs. (6) and (7) can be explicitly carried out from Eqs. (1) and (2). The compressibility and spin susceptibility sum rules can be satisfied by making at this point a long-wavelength approximation. This involves, as an instance, the replacement $\delta {\bar g}({\bf r}_1,{\bf r}')/\delta n({\bf r}) \rightarrow \delta({\bf r}_1-{\bf r})\partial {\bar g}(|{\bf r}-{\bf r}'|)/\partial n$. Some straightforward algebra leads to the expressions
\begin{equation}\label{kxcpiu}
K^{+}_{\mathrm  xc}({\bar n};r)=\frac{e^2}{r}\,\left[{\bar g}(r)-1+2\,{\bar n}\frac{\partial {\bar g}(r)}{\partial {\bar n}}+\frac{1}{2}\,{\bar n}^2\frac{\partial^2 {\bar g}(r)}{\partial {\bar n}^2}\right]
\end{equation}
and 
\begin{equation}\label{kxcmeno}
K^{-}_{\mathrm  xc}({\bar n};r)=\frac{e^2}{r}\,\left[\frac{1}{2}\,{\bar n}^2\left.\frac{\partial^2 {\bar g}(r)}{\partial {\bar m}^2}\right|_{{\bar m}=0}\right]~.
\end{equation}
The local field factors are thus given in this approximation by		                         
\begin{equation}\label{pdt+}
G_{+}(q)=\left(1+2{\bar n}\,\frac{\partial}{\partial {\bar n}}+\frac{1}{2}\, {\bar n}^2\frac{\partial^2}{\partial {\bar n}^2}\right)G(q)
\end{equation}
and
\begin{equation}\label{pdt-}
G_{-}(q)=\frac{1}{2}\,{\bar n}^2\,\left.\frac{\partial^2}{\partial {\bar m}^2}\, G(q)\right|_{{\bar m}=0}~,
\end{equation}
where we have defined
\begin{equation}\label{gstortino}
G(q)\equiv -\frac{1}{{\bar n} e^2}\int_{0}^{e^2}d \lambda \int \frac{d^{\, d} {\bf q}'}{(2\pi)^d}\,\frac{v_{q'}}{v_q}\,\left[S_{\lambda}(|{\bf q}-{\bf q}'|)-1\right]~.
\end{equation}
Here, $S_{\lambda}(q)$ is the static structure factor of the electron gas with a pair potential $v_\lambda(q)$  given by $4 \pi \lambda/q^2$ in $d=3$ and by $2 \pi \lambda/q$ in $d=2$. That is,		        \begin{equation}
S_{\lambda}(q)=1+{\bar n}\int d^d {\bf r}\left[g_{\lambda}(r)-1\right]\exp{(-i {\bf q}\cdot {\bf r})}~.
\end{equation}
Eqs. (10) - (12) are the main new results of the present work.	

Let us pause at this point to explicitly show that Eqs. (10) - (13) satisfy the two thermodynamic sum rules. By taking the long-wavelength limit $q \rightarrow 0$ inside the integral in Eq. (12) we find
\begin{equation}\label{step_0}
\lim_{q \rightarrow 0} G(q)=-\frac{2 \varepsilon_{xc}({\bar n},{\bar m})}{{\bar n}\, v_q}
\end{equation}
where
\begin{equation}
\varepsilon_{xc}({\bar n},{\bar m})=\frac{1}{2 e^2}\,\int_{0}^{e^2}d \lambda \int \frac{d^{\, d} {\bf q}'}{(2\pi)^d}\,v_{q'}\,
\left[S_{\lambda}(q')-1\right]
\end{equation}
is the exchange-correlation energy in the homogeneous, spin-polarized electron gas. Since $\varepsilon_{xc}$ is equal to the difference $\varepsilon_{gs}-\varepsilon_0$ between the ground-state energies of the real and of the non-interacting electron gas, it follows from Eqs.~(\ref{pdt+}) and (\ref{step_0}) that
\begin{equation}\label{compsr}
\lim_{q \rightarrow 0} G_{+}(q)=\frac{1}{{\bar n}^2\,v_q}\left(\frac{1}{K_{0}}-\frac{1}{K}\right)
\end{equation}
$K$ and $K_0$ being the corresponding compressibilities. Similarly, from Eqs. (11) and (14) we get
\begin{equation}
\lim_{q \rightarrow 0} G_{-}(q)=-\frac{1}{{\bar n}v_q}\,\left.\frac{\partial^2 \varepsilon_{\mathrm xc}({\bar n},\xi)}{\partial \xi^2}\right|_{\xi=0}
\end{equation}
with $\xi={\bar m}/{\bar n}$ and hence
\begin{equation}\label{sssr}
\lim_{q \rightarrow 0} G_{-}(q)=\frac{\mu^2_{B}}{v_{q}}\left(\frac{1}{\chi_P}-\frac{1}{\chi_s}\right)
\end{equation}
where $\mu_B$ is the Bohr magneton, $\chi_P$ the Pauli susceptibility and $\chi_s$ the spin susceptibility of the interacting electron gas. Equations (16) and (18) are the correct relations between the long-wavelength limit of the static response of the paramagnetic electron gas and the second derivatives of its ground state energy [12].	

Returning to Eqs. (10) - (12), they evidently need supplementing by a method for the evaluation of the electron-pair structure. Such a calculation can be carried out self-consistently by one of two alternative routes: (i) by using the static linear response functions to construct effective spin-dependent electron-electron interactions that yield the electron-pair distribution functions through the solution of an electron-electron scattering problem [13 - 15]; or (ii) by constructing dynamic susceptibilities from the static local-field factors and imposing that the fluctuation-dissipation formula relating the electron-pair distribution function to the van Hove dynamic structure factor is satisfied [16]. The former method has already been tested to some extent and shown to yield accurate results for quantum plasmas over a physically significant range of coupling strength [17]. In the following we present and discuss the second method mentioned above, in regard to the dielectric response and charge-charge correlations for electrons in $d = 3$ and $d = 2$. The study of spin-spin correlations requires entering the spin-polarized state, as already mentioned, and is left for future work.	

In brief, following the work of Singwi et al. [16] we write an approximate expression for the dynamic dielectric susceptibility $\chi_{c}(q,\omega)$ from the static local-field factor $G_{+}(q)$ as		     
\begin{equation}
\chi_{c}(q,\omega)=\frac{\chi_0(q,\omega)}{1-v_q[1-G_{+}(q)]\chi_0(q,\omega)}
\end{equation}				
where $\chi_0(q,\omega)$ is the dynamic (Lindhard-Stern) susceptibility of the ideal Fermi gas. The electron-pair structure factor can then be calculated as		                   
\begin{equation}
S(q)=-\frac{1}{\pi {\bar n}}\int_{0}^{\infty}d\omega~ {\rm Im} \chi_{c}(q,\omega)~.
\end{equation}
Self-consistence between pair structure and dielectric response is ensured by the fact that the local-field factor to be used in Eq. (19) depends on the pair structure according to Eqs. (10) and (12). Of course, Eqs. (19) and (20) hold for each value of $\lambda$ and must be solved together with Eqs. (10) and (12) up to the physical value of the coupling strength in the system.	

The results of this calculation for a weakly coupled electron gas in $d = 3$ are shown in Figures 1 and 2. It is seen from Figure 1 that the local-field factor  $G_+(q)$ obtained from the present theoretical approach is quite close to the QMC data of Moroni et al. [7] over a substantial range of values for $q/k_F$. Important deviations from these data are found only at high momenta $q > 2 k_F$, where the details of the dielectric response have little relevance for most practical purposes: in fact, in our approach the asymptotic value of $G(q)$ at large momenta is a constant determined by the quantity $1 -{\bar g}(r=0)$, rather than showing a leading term of order $q^2$. It is also seen from Figure 1 that our self-consistent results are in good agreement with the parametrized expression of $G_+(q)$ given by Ichimaru and Utsumi (IU) [18], who combined QMC data on the compressibility with a ladder diagram calculation of  $g(r=0)$     and a previous dielectric formulation by Utsumi and Ichimaru [11]. The peak of $G_+(q)$ for momenta approaching $2 k_F$, which in their formulation arose from a logarithmic term associated with exchange, comes in our approach from a logarithmic factor remaining inside the integral in Eq. (12) after angular integration in $d = 3$. This peak is absent in the so-called STLS dielectric theory of Singwi et al. [16], as can again be seen from Figure 1.	

Our results for the pair distribution function $g(r)$ at full self-consistence in $d = 3$ are reported in Figure 2, in comparison with QMC data kindly provided by P. Ballone (unpublished) and with the results of the STLS theory. Evidently, the present results for the electron-pair structure are excellent in this weak-coupling regime. With increasing coupling strength the above-mentioned peak in $G_+(q)$, however, introduces unwanted oscillations in $g(r)$. It appears that as the coupling increases the frequency dependence of the local-field factors [19] may become important in the calculation of pair correlations via the fluctuation-dissipation theorem. Much more satisfactory results are instead obtained by calculating these correlations through the solution of an electron-electron scattering problem with effective scattering potentials embodying the many-body effects from the static local-field factors [17].

Finally, Figure 3 reports similar results for the local-field factor $G_+(q)$ in a weakly coupled electron gas in $d = 2$. In this case satisfying the compressibility sum rule ensures that the local-field factor from QMC data [12] is reproduced almost up to $q \simeq 2 k_F$. Again a peak is found to be present in $G_+(q)$ for $q \simeq 2 k_F$, which is absent in an STLS formulation.	

In summary, we have proposed a self-consistent theoretical approach to the evaluation of the local-field factors entering the dielectric and longitudinal-spin response of an interacting electron gas and given some examples of the results that it predicts for the dielectric response and the charge-charge correlations at weak coupling. In the present formulation the form of the local-field factors is very accurate in the most important range of momenta where the thermodynamic sum rules dominate the physical behavior, but may still be too sensitive to exchange effects at momenta of order $2 k_F$. 
For this reason, structural self-consistence appears to be best enforced at strong coupling through the use of the local-field factors in constructing effective many-body scattering potentials. 
\vspace{0.5 cm}

{\bf Ackowledgements}
\vspace{0.2 cm}

This work was partially funded by MURST under the PRIN2001 Initiative. We wish to thank the Condensed Matter Group of the Abdus Salam International Centre for Theoretical Physics in Trieste for their hospitality during the final stages of this work. We are indebted with Professor P. Ballone for providing us with the QMC data reported in Figure 2.
\newpage

{\bf References}
\vspace{0.2 cm}

[1]  See, e.g. D. Ceperley, Nature 397 (1999) 386. 

[2]  R.M. Dreizler, E.K.U. Gross, {\it Density Functional Theory, An Approach to the Quantum Many-Body Problem} 

\hspace{0.35 cm} (Springer, Berlin 1990). 
 
[3]  S. Yarlagadda, G.F. Giuliani, Phys. Rev. B 40 (1988) 5432 and B 49 (1994) 14188.  

[4]  V.M. Pudalov, M.E. Gershenson, H. Kojima, N. Butch, E.M. Dizhur, G. Brunthaler, 	A. Prinz, G. Bauer, 

\hspace{0.35 cm} Phys. Rev. Lett. 88 (2002) 196404. 
 
[5]  D.M. Ceperley, B.J. Alder, Phys. Rev. Lett. 45 (1980) 566. 

[6]  B. Tanatar, D.M. Ceperley, Phys. Rev. B 39 (1989) 5005.  

[7]  S. Moroni, D.M. Ceperley, G. Senatore, Phys. Rev. Lett. 75 (1995) 689.  

[8]  G. Ortiz, M. Harris, P. Ballone, Phys. Rev. Lett. 82 (1999) 5317. 

[9]  D. Varsano, S. Moroni, G. Senatore, Europhys. Lett. 53 (2001) 348.

[10] P. Vashishta, K.S. Singwi, Phys. Rev. B 6 (1972) 875 and 4883. 

[11] S. Ichimaru, Rev. Mod. Phys. 54 (1982) 1017. 

[12] B. Davoudi, M. Polini, G.F. Giuliani, M.P. Tosi, Phys. Rev. B 64 (2001) 153101 and 	233110. 

[13] A.W. Overhauser, Can. J. Phys. 73 (1995) 683. 

[14] P. Gori-Giorgi, J.P. Perdew, Phys. Rev. B 64 (2001)155102. 

[15] B. Davoudi, M. Polini, R. Asgari, M.P. Tosi, Phys. Rev. B 66 (2002) 075110. 

[16] K.S. Singwi, M.P. Tosi, R.H. Land. A. Sj\"{o}lander, Phys. Rev. 176 (1968) 589. 

[17] B. Davoudi, M. Polini, R. Asgari, M.P. Tosi, cond-mat/0206456. 

[18] S. Ichimaru, K. Utsumi, Phys. Rev. B 24 (1981) 7385. 

[19] R. Nifos\`\i, S. Conti, M.P. Tosi, Phys. Rev. B 58 (1998) 12758. 
\newpage
\begin{figure}
\centerline{\mbox{\psfig{figure=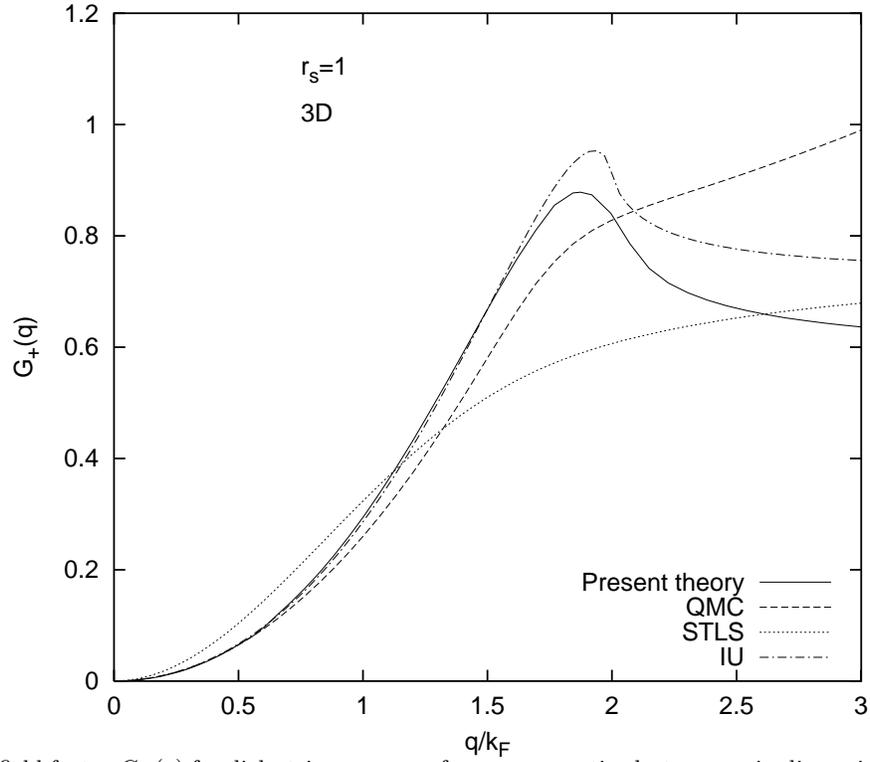, angle =0, width =12 cm}}} 
\caption{The local-field factor $G_+(q)$ for dielectric response of a paramagnetic electron gas in dimensionality $d=3$ at coupling strength $r_s\equiv(4 \pi {\bar n} a^3_B/3)^{-1/3}=1$, as a function of reduced momentum $q/k_F$.}
\label{Fig1}
\end{figure}
\newpage

\begin{figure}
\centerline{\mbox{\psfig{figure=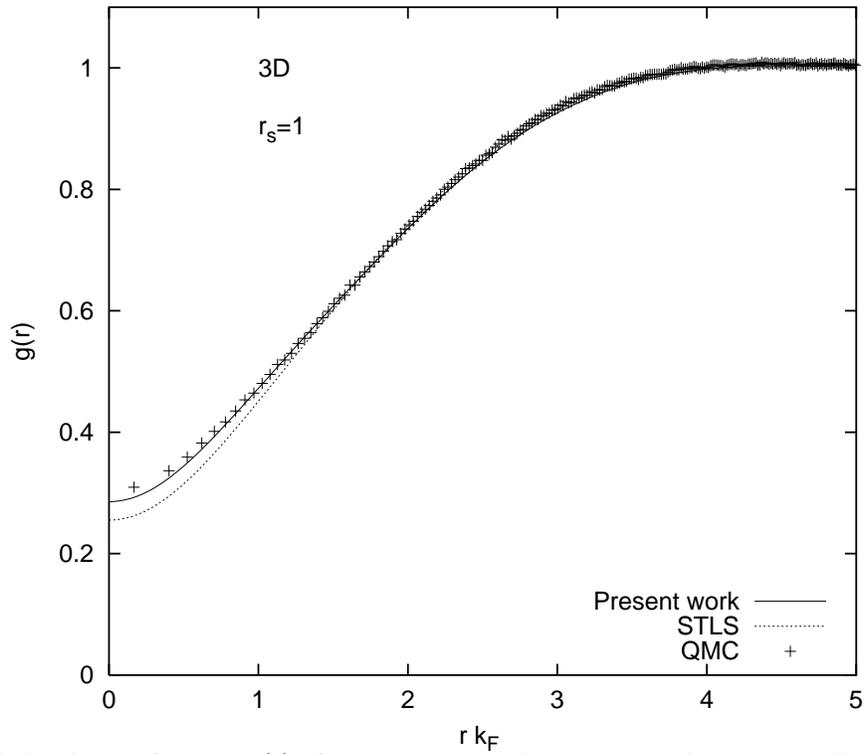, angle =0, width =12 cm}}} 
\caption{
The radial distribution function $g(r)$  of a paramagnetic electron gas in $d=3$ at coupling strength $r_s=1$, as a function of reduced distance $r k_F$.
}
\label{Fig2}
\end{figure}
\newpage

\begin{figure}
\centerline{\mbox{\psfig{figure=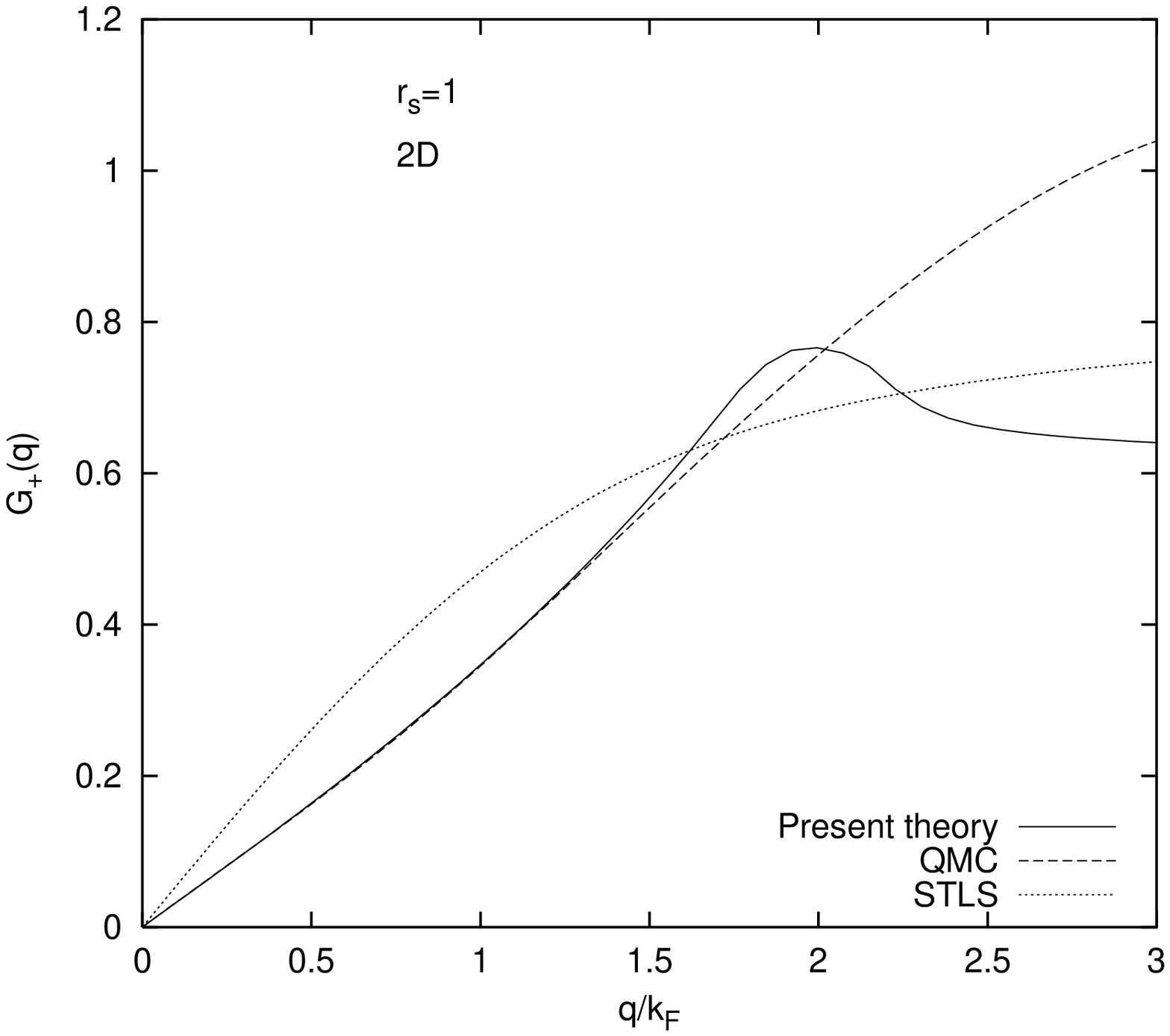, angle =0, width =12 cm}}} 
\caption{ 
The local-field factor $G_+(q)$ for dielectric response of a paramagnetic electron gas in $d=2$ at coupling strength $r_s\equiv ( \pi {\bar n}a^2_B)^{-1/2}$, as a function of reduced momentum $q/k_F$.
}
\label{Fig3}
\end{figure}
\end{document}